\documentclass[a4paper,11pt]{article}
\pdfoutput=1 % if your are submitting a pdflatex (i.e. if you have
\usepackage{jheppub} % for details on the use of the package, please
\usepackage[T1]{fontenc} % if needed

\usepackage[latin1]{inputenc}
\usepackage{graphicx}
\usepackage{enumitem}
\usepackage{color}

\usepackage{amsmath}
\usepackage{epsfig}
\usepackage{amsfonts}
\usepackage{amssymb}
\usepackage{cancel}
\usepackage{url}

\title{A basis of dimension-eight operators for anomalous neutral triple gauge boson interactions}

\author{Celine Degrande}
\affiliation{Department of Physics, University of Illinois at Urbana-Champaign\\
1110 W. Green Street, Urbana, IL 61801, USA          }

\emailAdd{cdegrand@illinois.edu}
\abstract{Four independent dimension-eight operators give rise to anomalous neutral triple gauge boson interactions, one CP-even and three CP-odd.  Only the CP-even operator interferes with the Standard Model for the production of a pair of on-shell neutral bosons. However, the effects are found to be tiny due mainly to the mismatch of the  Z boson polarization  between the productions from the SM and the new operator. }

\begin{document}

\maketitle

\section{Introduction}

The recent discovery of the Higgs boson has increased the confidence in the validity of the Standard Model (SM). On the other hand, the  remaining issues of the SM like the absence of a dark matter candidate claim for new physics. This dilemma can only be solved experimentally by either directly searching for new particles or by looking for deviations from the SM predictions. In this article, we use the well motivated effective field theory (EFT) approach to pin down the expected first deviations from heavy new physics on the neutral triple gauge couplings (nTGC).  \\
Anomalous neutral gauge couplings have been actively searched for at LEP~\cite{Acciarri:1999ug,Abbiendi:2003va,Alcaraz:2006mx}, at the Tevatron~\cite{Abazov:2007ad,Aaltonen:2011zc} and at the LHC~\cite{Chatrchyan:2012sga,Aad:2012mr}. The constraints are given following the parametrization of the anomalous vertices for the neutral gauge bosons  \cite{Gaemers:1978hg,Hagiwara:1986vm,Gounaris:2000dn,Gounaris:2000tb}
\begin{eqnarray}
i e \Gamma^{\alpha \beta \mu}_{ZZ V} (\text{q}_1, \text{q}_2, \text{q}_3)
&=& \frac{-e (\text{q}_3^2-m_V^2)}{M^2_Z}
\left [ f_4^V (\text{q}_3^\alpha g^{\mu \beta}+\text{q}_3^\beta g^{\mu \alpha})
-f_5^V \epsilon^{\mu \alpha \beta \rho}(\text{q}_1-\text{q}_2)_\rho \right ]
~, \label{eq:anoZZ} \\
i e\Gamma^{\alpha \beta \mu}_{Z\gamma V} (\text{q}_1, \text{q}_2, \text{q}_3)
&=& \frac{ -e (\text{q}_3^2-m_V^2)}{M_Z^2}
\Bigg \{ h_1^V (\text{q}_2^\mu g^{\alpha \beta}-\text{q}_2^\alpha g^{\mu \beta} )
+ \frac{h_2^V}{M_Z^2} \text{q}_3^\alpha [ (\text{q}_3\text{q}_2) g^{\mu \beta}- \text{q}_2^\mu \text{q}_3^\beta ]
\nonumber \\
&-& h_3^V \epsilon^{\mu \alpha \beta \rho} q_{2\rho}
~-~\frac{h_4^V}{M_Z^2} \text{q}_3^\alpha \epsilon^{\mu \beta \rho
\sigma}\text{q}_{3\rho} q_{2\sigma} \Bigg \}~  \label{eq:anoZA}
\end{eqnarray}
where V is a photon or a Z boson and is off-shell while the two other bosons are on-shell. The parametrization of those vertices has been extended for off-shell bosons in ref.~\cite{Gounaris:2000dn}. So far, the size of the $f^V_i$ and $h^V_i$ coefficients is unknown. Since the vertices contain three momenta, they are expected to arise from at least dimension-six operators. Therefore the contribution from heavy new physics is expected to be suppressed by at least the square of the ratio of the weak and the new physics scale. They have be computed or estimated for some extensions of the SM~\cite{Gounaris:2000dn,Ellison:1998uy}. Alternatively, their size as well as their dependence in a smaller number of parameters can be obtained for any heavy new physics model using EFT~\cite{Degrande:2012wf}. 
 As a matter of fact, any extension the SM can be parametrized at low energy by the effective Lagrangian
\begin{equation}
 \mathcal{L}=\mathcal{L}_{SM}+\sum_{d>4}\sum_i \frac{C_i}{\Lambda^{d-4}} \mathcal{O}_i^d \label{eq:Left}
\end{equation}
where $d$ is the dimension of the operator $\mathcal{O}_i^d$ and $\Lambda$ is the new physics scale. The new operators are built out of the SM fields and respect its gauge symmetries. In the limit $\Lambda\to\infty$, this Lagrangian tends to the SM one. At energies well below $\Lambda$, only the finite set of operators with the lowest dimension are relevant. Therefore the Lagrangian~\eqref{eq:Left} is predictive even if the coefficients $C_i$ are kept as free parameters and can be used to search for heavy new physics in a model independent way. \\
Since no dimension-six operator induces nTGC~\cite{Buchmuller:1985jz,Grzadkowski:2010es}, the effects of heavy new physics on nTGC is expected to be rather small. Whether or not they can be observed or constrained depends on the presence of dimension-eight operators with nTGC and the size of their contributions. 
Two dimension-eight operators invariant under the SM gauge group have been given in \cite{Gounaris:2001mw}. Those operators contain only the neutral gauge bosons and the Higgs field but they can be transformed using the equations of motion to operators without nTGC. 
In the following, we allow the operators to contain the charged gauge bosons as well since many dimension-six and dimension-eight operators can change they interactions and may cancel the effects of the operators with nTGC.\\
The list of independent operators is derived in section~\ref{sec:op} while their effects on neutral diboson production are discussed in section~\ref{sec:pheno}. Finally, a summary is given in the last section.

\section{The operators}\label{sec:op}

In this section, we go through all the possible dimension-eight operators invariant under the SM gauge group with neutral triple gauge vertices to find a set of independent operators. Namely, we are using 
\begin{enumerate}[label=\Roman{*}., ref=\Roman{*}]
\item The Higgs field equation of motion \label{it:eomH}
\begin{eqnarray}
 D_\mu D^\mu H& =& \mu^2 H -2 \lambda \left(H^\dagger H\right) H+ \text{fermion densities};
\end{eqnarray}
\item The W field equation of motion  \label{it:eomF}
\begin{eqnarray}
 D^\mu W^I_{\nu\mu} &=& i g \left(H^\dagger\sigma^I D_\nu H-D_\nu H^\dagger\sigma^I  H\right) + \text{fermion currents}
\end{eqnarray}
and the B field equation of motion
\begin{eqnarray}
 D^\mu B_{\nu\mu} &=& i \frac{g'}{2} \left(H^\dagger D_\nu H-D_\nu H^\dagger  H\right) + \text{fermion currents};
\end{eqnarray}
\item Bianchi identities  \label{it:bianchi}
\begin{eqnarray}
 D^\mu \widetilde{W}^I_{\nu\mu} = 0\qquad \text{and} \qquad
 D^\mu \widetilde{B}_{\nu\mu} =0;
\end{eqnarray}
\item Jacobi identity \label{it:jacobi}
\begin{equation}
 D_\mu F_{\nu\rho}+D_\nu F_{\rho\mu}+D_\rho F_{\mu\nu}=0
\end{equation}
where $F_{\nu\rho}$ denotes either $W_{\nu\rho}$ or $B_{\nu\rho}$ here and in the remaining of this paper;
\item The commutator of two covariant derivatives is a linear combination of strength field tensors; \label{it:comu}
\item Integration by part.\label{it:ibp}
\end{enumerate}
Rules \ref{it:eomH} and \ref{it:eomF} allow to transform an operator to operators with more Higgs fields or fermion fields or dimension-six operators with at least two Higgs fields. No dimension-six with two Higgs fields have nTGC even without using the equations of motion. As a matter of fact, only two independent (using rule~\ref{it:ibp} only) operators with four derivatives and two Higgs fields cannot be written as operators with at least one strength tensor,
\begin{equation}
\left(\left\{D^\mu, D^\nu\right\} H\right)^\dagger \left(\left\{D_\mu, D_\nu\right\} H\right)\quad \text{and}\quad \left(\left\{D^\mu, D_\mu\right\} H\right)^\dagger\left(\left\{D^\nu, D_\nu\right\} H\right).
\end{equation}
Since their parts with three neutral gauge bosons have an imaginary coefficient they cancel each other in those hermitian operators. If there is a strength tensor, the two derivatives can be chosen to be applied on the Higgs fields only. If they are acting on two different fields, they both give a Z and $Z^\mu Z^\nu$ gives zero when contracted with a strength tensor. If they are applied on the same field, they can be replaced by strength tensors and there is again no nTGC. 
Consequently, we will start with the operators with no Higgs and use those two rules to remove some operators with a low number of Higgs fields in favor of operators with more Higgs fields. 
Similarly, rule \ref{it:comu} allows to change the order of the covariant derivatives while generating operators with more strength tensors and less covariant derivatives. In particular, the operator is replaced by operators with more Higgs fields or more strength tensors
\begin{enumerate}[label=\Roman{*}., ref=\Roman{*}]
    \setcounter{enumi}{6}
\item  if two covariant derivatives acting on a Higgs fields are contracted together because they can be brought close to the Higgs fields using rule~\ref{it:comu} such that rule \ref{it:eomH} can be used or \label{it:H}
\item if a covariant derivative is contracted with the strength (dual) tensor on which it is acting because it can be moved with the help of rule~\ref{it:comu} such that rule \ref{it:eomF} (\ref{it:bianchi}) can be applied.\label{it:F}
\end{enumerate}
Using the equation of motion in this way, we only keep the operators with nTGC that are not equivalent to operators without nTGC (See for example Ref.~\cite{Cata:2013sva} for operators involving fermions currents). 
Finally, we assume that the gauge bosons are produced by the collision of massless fermions and they are produced either on-shell or decay into massless fermions consistently with the traditional searches for nTGC. Therefore operators with vertices proportional to the scalar product of the polarization vector and the momentum of the same boson  can be removed as well.

\subsection{Without the Higgs field}\label{sec:noH}

Operators with only one strength tensor or with two strength tensors can be removed using rules \ref{it:F} and \ref{it:ibp} as long as the two Lorentz indices of the strength tensors are not contracted together. Thoses particular cases can be removed using rule \ref{it:jacobi}.

Only the normal derivatives in the covariant derivatives can contribute to TGC in the operators with three strength tensors and two covariant derivatives. Furthermore, there is no neutral TGC if the three strength tensors are $SU(2)$ triplet contracted with $\epsilon^{IJK}$ and therefore there is at least one $B_{\mu\nu}$. This $B_{\mu\nu}$ can always be chosen without covariant derivatives acting on it by \ref{it:ibp}. Consequently, operators with two derivatives and three strength tensors can be written as
\begin{equation}
B_{\alpha\beta} X^{\alpha\beta}
\end{equation} 
where $X^{\alpha\beta}$ is an antisymmetric tensor made of two covariant derivatives and two identical strength tensors.
The covariant derivatives can be forbidden to have the same Lorentz index as the strength field on which they are applied thanks to rule \ref{it:F}. The cases where the two derivatives are adjacent and contracted together can be transformed into operators with one more strength tensors or with a current by using Jacobi identity.
Consequently, the various possible expression for $X^{\alpha\beta}$ when it contains two $W$ strength tensors  can be written as
\begin{eqnarray}
X^{\alpha\beta}&=&\left\langle D^\alpha D^\beta W^{\mu\nu} W_{\mu\nu}\right\rangle,\label{d201}\\
X^{\alpha\beta}&=&\left\langle D^\alpha D^\nu W^{\mu\beta} W_{\mu\nu}\right\rangle,\label{d202}\\
X^{\alpha\beta}&=&\left\langle D^\mu D^\nu W^{\alpha\beta} W_{\mu\nu}\right\rangle,\label{d203}\\
X^{\alpha\beta}&=&\left\langle D^\nu W^{\mu\beta} D^\mu W_{\alpha\nu}\right\rangle,\label{d113}\\
X^{\alpha\beta}&=&\left\langle D^\mu W^{\nu\beta} D^\mu W_{\alpha\nu}\right\rangle,\label{d114}\\
X^{\alpha\beta}&=&\left\langle D^\mu W^{\nu\beta} D^\alpha W_{\mu\nu}\right\rangle,\label{d115}\\
X^{\alpha\beta}&=&\left\langle D^\beta W^{\mu\nu} D^\alpha W_{\mu\nu}\right\rangle\label{d111}
\end{eqnarray}
where  $\left\langle \dots\right\rangle$ denotes the trace, or equivalent operators up to a sign or the exchange of $\alpha$ and $\beta$.The operators in Eqs~\eqref{d113}, \eqref{d114} and  \eqref{d111} are symmetric. In Eqs.~\eqref{d201} and  \eqref{d203}, the derivatives can be replaced by a strength tensor. The operators in Eqs.~\eqref{d202} and \eqref{d115} give equivalent operators when contracted with $B_{\alpha\beta}$  up to operators with a current through integration by part. The operator in Eq.~\eqref{d115} is related to the operators in Eqs.~\eqref{d113} and \eqref{d114} through jacobi identity. The arguments remain if the $W$ strength tensor are replaced by $B$ strength tensors or if $B_{\alpha\beta}$ is replaced by its dual tensor. Only the case with one dual $W$ strength tensor need to be further discussed. The operators from Eqs.~\eqref{d201} and  \eqref{d203} can still be replaced by operators with four strength tensors and operators from Eq.~\eqref{d202} can still be replaced by the one from Eq.\eqref{d115} independently of which $W$ got the tilde. The operators from Eq.~\eqref{d114} are equivalent to a linear combination of operators with two  derivatives contracted together acting on one of the strength tensors including $B_{\alpha\beta}$. The operators from Eq.~\eqref{d111} are equivalent to the operators from Eq.~\eqref{d201} and some operators with a current by \ref{it:ibp}. The derivative on the $W$ strength tensor on the operators from Eq.~\eqref{d113} can be moved to the two other strength tensors by \ref{it:ibp}. The term with this derivative acting on the $B$ strength tensor can be transformed into a operator with a dual $B$ strength tensor using
\begin{equation}
D_\mu B_{\alpha\beta} \widetilde{W}^{\mu\beta} =\frac{1}{2} D_\alpha \widetilde{B}^{\mu\beta}{W}_{\mu\beta}.\label{eq:jacotilde}
\end{equation}
This relation is obtained by multiplying Jacobi identity by the dual tensor. The remaining piece is removed by \ref{it:F}.
The operators in Eq.~\eqref{d115} with one dual strength tensor are still related to the operators from Eqs.~\eqref{d113} and \eqref{d114} through jacobi identity.
To sum up, there are no operators with three strength tensors and two covariant derivatives that contribute to nTGC and cannot be written as operators with two or more Higgs fields or  more strength tensors.

Finally, the operators with four strength tensors induce vertices with at least four fields and are therefore irrelevant.

\subsection{With two Higgs fields}\label{sec:2h}

In this section, we will remove the derivatives on the $H^\dagger$ field by integration by part. The operators with six covariant derivatives can be replaced by operators with at least one strength tensor or by operators with four Higgs fields using rule \ref{it:H}.

If there is one strength tensor and four derivatives, exactly one derivative should be applied to the strength tensor due to rules \ref{it:H} and \ref{it:F}. Those operators should therefore be written as
\begin{equation}
H^\dagger D_\rho F_{\mu\nu}D^\mu D^\rho D^\nu H
\end{equation}
or with a different order of the Lorentz indices for the derivatives acting on the Higgs field. If the derivatives carrying the same index as the strength tensor are adjacent, they can be replaced by a sum of strength tensors. If they are not, they can be brought again close to each other using the strength tensors. Consequently, we can forgot also the operators with four derivatives.

Operators with two derivatives and two strength tensors can be written as
\begin{eqnarray}
&H^\dagger  F_{\mu\nu}F^{\mu\rho} D_\rho D^\nu H&\label{eq:2H1}\\
&H^\dagger  D_\rho F_{\mu\nu}F^{\mu\rho} D^\nu H&\label{eq:2H2}\\
&H^\dagger  D_\rho F_{\mu\nu}D^\nu F^{\mu\rho}   H.\label{eq:2H4}&
\end{eqnarray}
The cases with the indices of the two derivatives summed together have been removed using rules \ref{it:eomH} and \ref{it:jacobi}.
The operators in Eq.~\eqref{eq:2H2} can be transformed into the previous operators by integration by part and one operator with a derivative on each Higgs fields which contains vertices with at least four vector bosons and eventually an operator with an extra current. Moreover, the two derivatives can be replaced here by they anticommutator as the commutator will be treated later. The operator in Eq.~\eqref{eq:2H4} do not contain nTGC because the indices of three vector fields can only be contracted with $\epsilon^{IJK}$. Three operators with nTGC remain after replacing $F_{\mu\nu}$ by $B_{\mu\nu}$ or $W_{\mu\nu}$,
\begin{eqnarray}
\mathcal{O}_{BW}&=&i\, H^\dagger   B_{\mu\nu}W^{\mu\rho} \left\{D_\rho,D^\nu\right\} H,\label{eq:obw}\\
\mathcal{O}_{WW}&=&i\, H^\dagger  W_{\mu\nu}W^{\mu\rho} \left\{D_\rho,D^\nu\right\} H,\label{eq:oww}\\
\mathcal{O}_{BB}&=&i\, H^\dagger  B_{\mu\nu}B^{\mu\rho} \left\{D_\rho,D^\nu\right\} H.\label{eq:obb}
\end{eqnarray}
Four extra operators with nTGC can be obtained with the dual strength tensors,
\begin{eqnarray}
\mathcal{O}_{\widetilde{B}W}&=&i\, H^\dagger  \widetilde{B}_{\mu\nu}W^{\mu\rho} \left\{D_\rho,D^\nu\right\} H,\label{eq:obtw}\\
\mathcal{O}_{B\widetilde{W}}&=& i\, H^\dagger  B^{\mu\nu}\widetilde{W}_{\mu\rho} \left\{D_\rho,D^\nu\right\} H,\label{eq:obwt}\\
\mathcal{O}_{\widetilde{W}W}&=& i\, H^\dagger  \widetilde{W}_{\mu\nu}W^{\mu\rho} \left\{D_\rho,D^\nu\right\} H,\label{eq:owwt}\\
\mathcal{O}_{\widetilde{B}B}&=&i\, H^\dagger  \widetilde{B}_{\mu\nu}B^{\mu\rho} \left\{D_\rho,D^\nu\right\} H.\label{eq:obbt}
\end{eqnarray}
The identity,
\begin{equation}
B_{\mu\nu}\widetilde{B}^{\nu\rho}=-\frac{1}{4}\delta_\mu^\rho B_{\nu\sigma}\widetilde{B}^{\nu\sigma},\label{eq:fftilde}
\end{equation}
and its equivalent for $W$ transform the two last operators to operators with more Higgs or dimension-six operators. The operators from Eq.~\eqref{eq:obwt} is equivalent to the one from Eq.~\eqref{eq:obtw} up to operators with more currents, more strength tensors or with quartic gauge boson couplings only due to Eq.~\eqref{eq:jacotilde}.  

If there are only three strength tensors in addition to the two Higgs fields, the operator is antisymmetric under the exchange of the three strength tensors and no nTGC can be generated. There are also no operators with one dual strength tensor and two strength tensors due to Eq.~\eqref{eq:fftilde}.

\subsection{With four Higgs fields}\label{sec:4H}

If the four derivatives are applied each on one Higgs doublet, only quartic interactions are generated. Two derivatives should then be applied on one Higgs field and they should have different Lorentz indices due to the Higgs equation of motion. Only the $Z$ field can be selected with the Higgs vacuum expectation value. Consequently, the part of the operators with four Higgs contributing to the nTGC can be written as $Z^\mu Z^\nu \partial_\mu Z_\nu$ and have all the same nTCG vertex proportional to 
\begin{equation}
\sum_{perm(1,2,3)} \eta^{\mu_1\mu_2} p_3^{\mu_3}.
\end{equation} 
This vertex does not contribute when the vector bosons are on-shell or attached to a massless fermion line. 
There are also no operator with one or two strength tensors with nTGC.

\section{Phenomenology}\label{sec:pheno}

Once added to the Lagrangian with their hermitian conjugate, the four operators found in section~\ref{sec:2h} produces nTCG vertices independent of the imaginary part of their coefficients. Namely, none of the combinations $O_i-O_i^\dagger$ contains vertices with three neutral gauge boson. The Lagrangian for nTGC can therefore be written as 
\begin{equation}
\mathcal{L}^{nTGC}= \mathcal{L}_{SM} + \sum_{i} \frac{C_i}{\Lambda^4}\left(O_i+O_i^\dagger\right)
\end{equation}
where $i$ run over the label of the four operators from equations \eqref{eq:obw} to \eqref{eq:obtw}. In the definitions of the operators we use the following convention :
\begin{equation}
 D_\mu \equiv \partial_\mu - i \frac{g'}{2} B_\mu Y - i g_w W_\mu^i \sigma^i
\end{equation}
 and 
\begin{align}
W_{\mu\nu} & = \sigma^I (\partial_\mu W^I_\nu - \partial_\nu W^I_\mu
	+ g \epsilon_{IJK} W^J_\mu W^K_\nu )\\
B_{\mu \nu} & =  (\partial_\mu B_\nu - \partial_\nu B_\mu)
\end{align}
with $\left\langle \sigma^I \sigma^J\right\rangle=\delta^{IJ}/2$.

The CP-conserving anomalous couplings for the production of two on-shell Z bosons (see Eq.~\eqref{eq:anoZZ}) are given by
\begin{eqnarray}
f_5^Z&=&0\\
f_5^\gamma &=&\frac{  v^2 M_Z^2 }{4 c_w s_w} \frac{C_{\widetilde{B}W}}{\Lambda^4}
\end{eqnarray}
 and the CP-violating by
\begin{eqnarray}
f_4^Z &=&  \frac{ M_Z^2 v^2 \left(c_w{}^2 \frac{C_{WW}}{\Lambda ^4}+2 c_w s_w \frac{C_{BW}}{\Lambda ^4}+4 s_w{}^2
   \frac{C_{BB}}{\Lambda ^4}\right)}{2 c_w s_w}\\
f_4^\gamma &=&  -\frac{ M_Z^2 v^2 \left(-c_w s_w
  \frac{C_{WW}}{\Lambda ^4}+\frac{C_{BW}}{\Lambda ^4} \left(c_w{}^2-s_w{}^2\right)+4 c_w s_w \frac{C_{BB}}{\Lambda
   ^4}\right)}{4 c_w s_w}
\end{eqnarray}
 For one on-shell Z boson and one on-shell photon (see  Eq.~\eqref{eq:anoZA}), the CP conserving couplings are
\begin{eqnarray}
h_3^Z&=&\frac{  v^2 M_Z^2 }{4 c_w s_w} \frac{C_{\widetilde{B}W}}{\Lambda^4}\\
h_4^Z&=&  0\\
h_3^\gamma&=&0\\
h_4^\gamma&=&0\\
\end{eqnarray}
 and the CP-violating couplings are
\begin{eqnarray}
h_1^Z&=&\frac{ M_Z^2 v^2 \left(-c_w s_w \frac{C_{WW}}{\Lambda
   ^4}+\frac{C_{BW}}{\Lambda ^4} \left(c_w{}^2-s_w{}^2\right)+4 c_w s_w \frac{C_{BB}}{\Lambda ^4}\right)}{4 c_w
   s_w}\\
h_2^Z&=&0\\
h_1^\gamma&=&-\frac{
   M_Z^2 v^2 \left(s_w{}^2 \frac{C_{WW}}{\Lambda ^4}-2 c_w s_w \frac{C_{BW}}{\Lambda
   ^4}+4 c_w{}^2 \frac{C_{BB}}{\Lambda ^4}\right)}{4 c_w s_w}\\
h_2^\gamma&=&0.
\end{eqnarray}
The anomalous couplings for off-shell bosons are given in appendix~\ref{ap:ano}. Many couplings are zero because the CP-conserving $ZAA$  and $ZZZ$ vertices vanish. Those expressions implies two relations between the non-vanishing couplings, 
\begin{equation}
 f_5^\gamma=h_3^Z \qquad \text{and} \qquad h_1^Z=-f_4^\gamma.\label{eq:ACrel}
\end{equation}
$C_{\widetilde{B}W}/\Lambda^4$ can be constrained using the mearsurements of $ f_5^\gamma$ and $h_3^Z$. The limits on $ f_5^\gamma$~\cite{Chatrchyan:2012sga} give the strongest constraint, i.e.
\begin{equation}
 -47 \text{ TeV}^{-4}<\frac{C_{\widetilde{B}W}}{\Lambda^4}<47 \text{ TeV}^{-4}\label{eq:cbtwbound}
\end{equation}
at 95\% C.L. 
The coefficients on the CP-violating operators can be constrained from the measurements of $f_4^Z$,  $f_4^\gamma$~\cite{Chatrchyan:2012sga} and $h_1^\gamma$~\cite{Beringer:1900zz},
\begin{eqnarray}
 -47 \text{ TeV}^{-4}<&\frac{C_{BB}}{\Lambda^4}&<48 \text{ TeV}^{-4}\\
 -114 \text{ TeV}^{-4}<&\frac{C_{BW}}{\Lambda^4}&<113 \text{ TeV}^{-4}\\
 -99 \text{ TeV}^{-4}<&\frac{C_{WW}}{\Lambda^4}&<101 \text{ TeV}^{-4}.
\end{eqnarray}
Only the constraints on anomalous couplings obtained without form factor have been used to extract the limits on the coefficients of the dimension-eight operators. For order one coefficients, the new physics scale is constrained to be above a few hundred GeV. Such large coefficients require a strongly interacting new sector since the anomalous operators arise at the loop level in a UV complete theory.
Consequently, the coefficients are expected to be smaller in a weakly coupled model and the mass limit lower. For example, fermions with a non-zero axial coupling to the Z boson can generate the Levi-Civita tensor of the CP-conserving operator at one-loop. The contribution such a heavy fermion implies~\cite{Gounaris:2000dn}
\begin{equation}
\frac{C_{\widetilde{B}W} }{\Lambda^4} = \frac{e^2 Q g_A g_V}{2 \pi^2 c_w s_w M_F^2 v^2}
\end{equation}  
where $Q$, $g_A$, $g_V$, $M_F$ are respectively the charge, the axial and vector couplings to the Z boson and the mass of the fermion (see Eq.~42 in \cite{Gounaris:2000dn} for the couplings convention). The limit in Eq.~\eqref{eq:cbtwbound} forces the fermion mass to be of a few tens of GeV depending on its charge and couplings to the Z boson. It should be noted that the dependence in the inverse fermion mass in quadratic and not quartic as expected from the dimension of the operator. However, the fermion mass is related to the EWSB scale since the left and right-handed fermions have to transform differently under the weak gauge group to generate the axial coupling. It the new physics is due to new fermion or scalar multiplets, the CP-violating nTCG requires more than one loop contribution as their  interaction to the photon or the Z-boson do not break CP. The CP breaking term can come either from the interaction to the charged vector bosons or from the scalar potential. In these cases, the CP-violating coefficients can be estimated by
\begin{equation}
 C_i \sim \frac{e^2 g_{CP} g_{\cancel{CP}}}{(4\pi)^4}
\end{equation} 
where $ g_{CP}$ and $g_{\cancel{CP}}$ are CP-conserving and CP-violating couplings of the new sector. Again, the experimental bounds imply that the new physics scale as to be of the order a few tens of GeV. In both case, the limits are only rough estimates as the effective approach is not expected to be reliable for such low masses.

If the new physics is heavy, the largest new physics contribution to $f\bar{f}\to ZZ/ZA$ is expected from the interference between the SM and the dimension-eight operators,
\begin{equation}
\left|M\right|^2 = \underbrace{\left|M_{SM}\right|^2}_{\mathcal{O}\left(\Lambda^0\right)}+\underbrace{2\Re\left(M_{SM}M_{dim8}^*\right)}_{\mathcal{O}\left(\Lambda^{-4}\right)}+\underbrace{\dots}_{\mathcal{O}\left(\Lambda^{-6}\right)}+\underbrace{\left|M_{dim8}\right|^2+\dots}_{\mathcal{O}\left(\Lambda^{-8}\right)}+\mathcal{O}\left(\Lambda^{-10}\right).
\end{equation}
The first  and second dots represent the interference between the SM and the dimension-ten and dimension-twelve operators respectively. In the following, the given partial $\mathcal{O}\left(\Lambda^{-8}\right)$ contribution is the squared of the amplitude with dimension-eight operators. This term is given to understand the suppression of the interference but do not induce a leading contribution from the heavy new physics unless the interferences between the SM and the dimension-eight and dimension-ten operators are both strongly suppressed.  
Although the dimension-six operators do not induce nTGC at the tree-level, they can have a effect on nTGC at one-loop. Those contributions would be of the order $\frac{\alpha_{EM}}{4\pi} \frac{s}{\Lambda^2}$ while the tree-level contribution from the dimension-eight operators are of the order $\frac{s v^2}{\Lambda^4}$. Consequently, the contribution of the dimension-eight operators dominates the one-loop contribution of the dimension-six operator for 
\begin{equation}
\Lambda\lesssim \sqrt{\frac{4\pi}{\alpha_{EM}} s}\approx 10\text{ TeV}.
\end{equation}

The three first CP-odd operators do not contribute to the on-shell production of two neutral gauge bosons at the order $\Lambda^{-4}$. On the contrary, the  CP-even operator contributes to ZZ and ZA productions thanks to the $ZZA$ vertex. As a consequence, only the CP-even operator will be studied in the following.
%stop : add lambda>3s

\subsection{ZZ production}

The new operator only contributes to the production of one longitudinally polarized boson and one transversally polarized boson,
\begin{eqnarray}
\!\!\!\!\!\!&\!\!\!&\!\!\!\!\!\!2 \Re\left(M_{SM}M_{dim8}^* \right)(f\bar{f}\to Z Z)= 2 \Re\left(M_{SM}M_{dim8}^* \right)(f\bar{f}\to Z_T Z_L) =\nonumber\\
\!\!\!\!\!\!&\!\!\!&\!\!\!\!\!\! e^4  v^2 N_c Q\left(Q (Q+1) s_w^4-2 Q s_w^2
   T_3+T_3^2\right)\frac{C_{\widetilde{B}W}}{ \Lambda^4}\frac{  \left(4 M_Z^2-s\right)    s\left(2 \left(c_\theta^2+1\right) M_Z^2- s_\theta^2 s\right)}{2 c_w^3 s_w^3 
   \left(4 M_Z^4-4  s_\theta^2 M_Z^2 s + s_\theta^2 s^2\right)}\label{eq:intzz}
\end{eqnarray}
where the average over the spin is done but not over the color, $Q$ and $T_3$ are the fermion charge and weak isospin, $s$ the center of mass energy, $\theta$ the scattering angle, $N_c$ is the number of color and should be set to one for leptons. The dependence in $Q$ and $T_3$ is due to the fact that only the new physics diagram with the photon in the s-channel has an effect.  
By comparison, the SM contribution is
\begin{eqnarray}
\left|M_{SM}\right|^2 (f\bar{f}\to Z Z) &=& e^4  N_c (2 Q^4 s_w^8 - 4 Q^3 s_w^6 T_3 + 6 Q^2 s_w^4 T_3^2 - 4 Q s_w^2 T_3^3 + T_3^4)\nonumber\\&&\!\!\!\!\frac{ s \left(16 (1 + c_\theta^2) M_Z^6 - 4 (3 - 7 c_\theta^2 + 4 c_\theta^4) M_Z^4 s  - 8 s_\theta^2 c_\theta^2 M_Z^2 s^2  - (-1 + c_\theta^4) s^3\right) }{
c_w^4  s_w^4 \left(4 M_Z^4-4  s_\theta^2 M_Z^2 s + s_\theta^2 s^2\right)^2}\nonumber\\
\end{eqnarray}  
and becomes constant at high energy while the new physics contribution grows like $s$. Futhermore, the SM contribution to one longitudinally polarized boson and one transversally polarized boson,
\begin{eqnarray}
\left|M_{SM}\right|^2(f\bar{f}\to Z_L Z_T) &=& e^4  N_c (2 Q^4 s_w^8 - 4 Q^3 s_w^6 T_3 + 6 Q^2 s_w^4 T_3^2 - 4 Q s_w^2 T_3^3 + T_3^4)\nonumber\\&&\frac{2 s  \left(4 \left(c_\theta^2+1\right) M_Z^4-4  s_\theta^2 M_Z^2 s+s_\theta^4 s^2 \right)}{
c_w^4  s_w^4  \left(4 M_Z^4-4  s_\theta^2 M_Z^2 s + s_\theta^2 s^2\right)^2}
\end{eqnarray} 
goes like $1/s$ at high energy. As expected, the SM produces mainly transversally polarized bosons at high energy. Those analytical expressions also show that the NP and the SM contributions differ by their dependences in the scattering angle. In particular, the interference changes of sign for $c_\theta^2=\frac{s-2M_Z^2}{s+2M_Z^2}$.

Numerically, the interference between new operator and the SM is small. For example, the  the cross-section for an electron positron collider of 200 GeV is
\begin{equation}
\sigma (e\bar{e}\to Z Z)/fb = 1252 - 3.2\cdot 10^{-3}C_{\widetilde{B}W} \left(\frac{\text{1 TeV}}{\Lambda}\right)^4+\dots+1.4 \cdot10^{-4} C_{\widetilde{B}W}^2 \left(\frac{\text{1 TeV}}{\Lambda}\right)^8+\dots
\end{equation}
The $1/\Lambda^8$ term from the new physics amplitude squared should be neglected for phenomenological purpose. However, it has been kept only to show that the interference is highly suppressed as already mentioned. The fact that the new physics only contributes to one longitudinally polarized boson and one transversally polarized boson partially explains this suppression. As a matter of fact, the SM contribution to those polarizations is only one third of its total rate at this energy. If a cut on the scattering angle, $c_\theta^2<\frac{s-2M_Z^2}{s+2M_Z^2}$, is added,
\begin{equation}
\sigma_{\theta cut} (e\bar{e}\to Z_T Z_L)/fb = 233 - 2.6\cdot 10^{-2}C_{\widetilde{B}W} \left(\frac{\text{1 TeV}}{\Lambda}\right)^4+\dots+7.9 \cdot10^{-5} C_{\widetilde{B}W}^2 \left(\frac{\text{1 TeV}}{\Lambda}\right)^8+\dots
\end{equation}
the  interference suppression is further reduced. In fact,  the two parts of the interference with an opposite sign nearly cancel each other close to threshold. 
The new physics contribution is expected to be larger compared to the SM at higher energies but remains challenging since the new physics scale should also be pushed to higher energy for consistency. For a collision of 1 TeV, the new physics contribution is well below the fb for a new physics scale at a few TeV, i.e.
\begin{equation}
\sigma (e\bar{e}\to Z Z)/fb = 144 - 0.41 C_{\widetilde{B}W} \left(\frac{\text{1 TeV}}{\Lambda}\right)^4+\dots+7.1 C_{\widetilde{B}W}^2 \left(\frac{\text{1 TeV}}{\Lambda}\right)^8+\dots.
\end{equation}
The suppression of the interference remains at high energy but the cut in the scattering angle does not have such a large effect anymore,
\begin{equation}
\sigma_{\theta cut} (e\bar{e}\to Z_T Z_L)/fb = 0.59 - 0.44 C_{\widetilde{B}W} \left(\frac{\text{1 TeV}}{\Lambda}\right)^4+\dots+6.9 C_{\widetilde{B}W}^2 \left(\frac{\text{1 TeV}}{\Lambda}\right)^8+\dots
\end{equation}
The suppression is rather explained by the fact that the polarization from the SM contribution are mainly transverse far from threshold. Only slightly less than a percent of the SM production has one longitudinal Z boson at 1 TeV.
The remaining part of the suppression of the interference at both energies comes from the large phase difference between the SM and new physics amplitudes.
Since the interference between the SM and the dimension-eight operators is quite suppressed, the next order in the $1/\Lambda$ expansion may have a larger contribution for a new physics scale not very far from the energy probed by the experiment.

%\begin{eqnarray}
%2 \Re\left(M_{SM}M_{dim8}^* \right)(f\bar{f}\to Z_{T_\parallel} Z_L) &=& e^4  v^2 N_c Q\left(Q (Q+1) s_w^4-2 Q s_w^2 T_3+T_3^2\right)\nonumber\\&&\frac{C_{\widetilde{B}W}}{ \Lambda^4}\frac{  \left(4 M_Z^2-S\right)    S\left(2  M_Z^2-S s_\theta^2\right)}{2 c_w^3 s_w^3 \left(4 M_Z^4-4   M_Z^2 S s_\theta^2+S^2 s_\theta^2\right)}
%\end{eqnarray}

\subsection{AZ production}

The dimension-eight operator contributes both to the production of longitudinally and transversally polarized Z boson,
\begin{eqnarray}
2 \Re\left(M_{SM}M_{dim8}^* \right)(f\bar{f}\to A_T Z_L)/fb &=&e^4 N_c Q  v^2 \frac{C_{\widetilde{B}W}}{\Lambda ^4} \frac{ \left(2Q T_3 s_w{}^2-T_3^2\right)}{2 c_w{}^3
   s_w{}^3} s \\
2 \Re\left(M_{SM}M_{dim8}^* \right)(f\bar{f}\to A_T Z_T)/fb &=& e^4  N_c Q v^2 \frac{C_{\widetilde{B}W}}{\Lambda ^4} \frac{ \left(2Q T_3 s_w{}^2-T_3^2\right)}{2 c_w{}^3 s_w{}^3}M_Z^2 .
\end{eqnarray}
However, there are no new physics contributions to all the possible combination of transversally polarized Z boson. In fact, the polarization vectors of the Z boson and the photon have to be orthogonal to each other. At high energy, the production longitudinal Z boson dominates.   
On the contrary, the SM contribution, 
\begin{eqnarray}
\left|M_{SM}\right|^2(f\bar{f}\to A_T Z_L)/fb &=& 4 e^4  N_c Q^2 \left(2 Q^2 s_w{}^4-2 Q T_3 s_w{}^2+T_3^2\right)\frac{ M_Z^2 s}{c_w{}^2 s_w{}^2 \left(s-M_Z^2\right)^2}\nonumber\\\\
\left|M_{SM}\right|^2(f\bar{f}\to A_T Z_T)/fb &=& 2   e^4 N_c Q^2\left(2 Q^2 s_w{}^4-2 Q T_3 s_w{}^2+T_3^2\right)\frac{\left(c_\theta^2+1\right)  \left(M_Z^4+s^2\right) }{
  s_w^2c_w^2  s_\theta^2 \left(s-M_Z^2\right)^2},\nonumber\\
\end{eqnarray}
contains mainly transversally polarized Z bosons at high energy. Contrary to the new physics contribution, the SM one depends on the scattering angle. Namely, the bosons are produced mainly at small scattering angle in the SM. Finally, the dependence  in the center of mass energy is the same as for Z boson pair production for both the SM and the new operator.
 
 For the following cross-sections, we have only applied a cut on the $p_T$ of the photon, i.e. $p_T>10$ GeV, to remove the divergence.
 For a center of mass energy of 200 GeV, the new physics effects are of the order of a few fb if the new physics scale is at 600 GeV,
\begin{eqnarray}
\sigma(e\bar{e}\to A_T Z_L)/fb &=& 1364+0.383 C_{\widetilde{B}W} \left(\frac{\text{1 TeV}}{\Lambda}\right)^4+\dots\nonumber\\&&+2.11 \cdot10^{-3} C_{\widetilde{B}W}^2 \left(\frac{\text{1 TeV}}{\Lambda}\right)^8+\dots\\
\sigma(e\bar{e}\to A_T Z_T)/fb& =& 15620+7.96\cdot10^{-2} C_{\widetilde{B}W} \left(\frac{\text{1 TeV}}{\Lambda}\right)^4+\dots\nonumber\\&&+4.53 \cdot10^{-4} C_{\widetilde{B}W}^2 \left(\frac{\text{1 TeV}}{\Lambda}\right)^8+\dots
\end{eqnarray}
We remind once more that the $1/\Lambda^8$ terms are only given to understand the suppression of the interference.
At 1 TeV and for a new physics scale around 3 TeV, the effects of the new physics are slightly below 0.01 fb,
\begin{eqnarray}
\sigma(e\bar{e}\to A_T Z_L)/fb &=& 1.75+0.48 C_{\widetilde{B}W} \left(\frac{\text{1 TeV}}{\Lambda}\right)^4+\dots\nonumber\\&&+2.63 C_{\widetilde{B}W}^2 \left(\frac{\text{1 TeV}}{\Lambda}\right)^8+\dots \\
\sigma(e\bar{e}\to A_T Z_T)/fb &=& 866+4.02\cdot 10^{-3} C_{\widetilde{B}W} \left(\frac{\text{1 TeV}}{\Lambda}\right)^4+\dots\nonumber\\&&+2.17 \cdot10^{-2} C_{\widetilde{B}W}^2 \left(\frac{\text{1 TeV}}{\Lambda}\right)^8+\dots 
\end{eqnarray}
At that energy, the secondary longitudinal polarizations are at the level of the percent and the interference is highly suppressed. Like for ZZ production, there is a mismatch between the polarization states produced at high energy by the SM and by the new operator. On the one hand, it allows to disentangle the new physics contribution from the SM one. On the other hand, it suppress the interference. In fact, this mismatch is also present for charged bosons pair production from the SM and the dimension-six operators with the Higgs field~\cite{Degrande:2013mh}. Finally a cut on the scattering angle would improve the ratio of the new physics contribution over the SM one but will this time reduce the already small new physics contribution.\\
Analytical results presented above for both $ZZ$ and $ZA$ productions have been found in agreement for the total rate and the matrix element of one phase space point using FeynRules UFO output~\cite{Christensen:2008py,Degrande:2011ua}, ALOHA\cite{deAquino:2011ub} and Madgraph5~\cite{Alwall:2011uj}

\section{Conclusion}

If the new physics is heavier than the scale probed by the experiments, it can be described in a model independent way by an effective Lagrangian. Since no dimension-six operator induces nTGC, the first new physics contribution to those interactions comes from the dimension-eight operators.
We have found that only one CP-even and three CP-odd independent dimension-eight operators give rise to nTGC for on-shell bosons or bosons decaying to massless fermions. Since the CP even one only induces the ZZA anomalous vertex, only the $f_5^\gamma$ and $h^Z_3$ CP-even anomalous couplings do not vanish. Similarly, two of the CP-odd anomalous couplings vanish despite that the three anomalous vertex are generated by CP-odd dimension-eight operators. The two relations from Eq.~\eqref{eq:ACrel} further reduce the number independent parameters in the anomalous vertex approach.
Only the CP-even operator contributes to the production of a pair of on-shell neutral electroweak bosons. This operator affects both to ZZ and ZA productions. 
As expected, the contributions from the higher dimensional operator become larger compared to those of the SM at high energy. They also differ by the polarization of the produced bosons. At high energy, this operator produces almost exclusively one of the Z boson with a longitudinal polarization contrary to the SM.  The new physics contributions also differ from the SM ones by the dependence in the scattering angle. In particular, the interference between the SM and the dimension-eight operators change of sign at $c_\theta^2=\frac{s-2M_Z^2}{s+2M_Z^2}$ for ZZ production and the two pieces nearly cancel each other near threshold. The new physics contributions are therefore quite distinguishable from the pure SM contribution. Although we have only used the tree-level expression for the SM contribution, the SM behavior at high energy should not be modified by loop corrections.
Unfortunately, the difference between the SM and new physics contributions also implies that the interference is strongly suppressed.  Consequently, advanced analysis like matrix element method would be welcome to observe or constrain the coefficients of the effective operators if the new physics scale is at or above the TeV.  The present constraints on the anomalous couplings barely constrain the new physics scale to be above the weak scale.
Finally, the four dimension-eight operators are implemented in FeynRules model~\cite{FRNTGC} and can therefore be used in MadGraph5~\cite{MGNTGC}.

\acknowledgments

We are grateful for discussion with J.-M. G\'erard. We thanks S. Protopapadaki and D. Fontes for their careful checks of these results. This material is based upon work supported in part by the U. S. Department of Energy under Contract No. DE-FG02-13ER42001.

\appendix

\section{Translation to off-shell anomalous couplings}\label{ap:ano}

The contributions of the dimension-eight operators to the off-shell anomalous couplings~\cite{Gounaris:2000dn} are given by
\begin{eqnarray}
&f_1^{AAZ}=f_2^{AAZ}=f_3^{AAZ}=0&
%f_1^{AAZ}&=& \frac{i   e \left(\text{q}_3.\text{q}_3\right) v^2 \left(\frac{C_{\widetilde{B}W}}{\Lambda^4}+\frac{C_{B\widetilde{W}}}{\Lambda^4}\right)}{4}\\
%f_2^{AAZ}&=& \frac{i e \left(\text{q}_1.\text{q}_1-\text{q}_2.\text{q}_2\right) v^2   \left(\frac{C_{\widetilde{B}W}}{\Lambda ^4}+\frac{C_{B\widetilde{W}}}{\Lambda ^4}\right)}{4}\\
%f_3^{AAZ}&=& \frac{i e v^2 \left(\frac{C_{\widetilde{B}W}}{\Lambda ^4}+\frac{C_{B\widetilde{W}}}{\Lambda ^4}\right)}{2},
\end{eqnarray}

\begin{eqnarray}
f_1^{ZZA}&=& \frac{  v^2 \text{q}_3^2 }{4 c_w s_w} \frac{C_{\widetilde{B}W}}{\Lambda^4}\\
   f_2^{ZZA}&=& \frac{  v^2 \left(\text{q}_1^2-\text{q}_2^2\right) }{4 c_w s_w} \frac{C_{\widetilde{B}W}}{\Lambda^4}\\
   f_3^{ZZA}&=& 0%-\frac{i e v^2 \left(\frac{C_{\widetilde{B}W}}{\Lambda ^4} \left(2 c_w{}^2+s_w{}^2\right)-\frac{C_{B\widetilde{W}}}{\Lambda ^4} \left(c_w{}^2+2 s_w{}^2\right)\right)}{2 c_w s_w}
\end{eqnarray}
and
\begin{eqnarray}
&f_1^{ZZZ}=f_2^{ZZZ}=f_3^{ZZZ}=0&
\end{eqnarray}
for the CP-conserving couplings and by
\begin{eqnarray}
\widetilde{f}_1^{AAZ}&=& -\frac{
   \left(\text{q}_1^2-\text{q}_2^2\right) v^2 \left(s_w{}^2 \frac{C_{WW}}{\Lambda ^4}-2 c_w s_w \frac{C_{BW}}{\Lambda
   ^4}+4 c_w{}^2 \frac{C_{BB}}{\Lambda ^4}\right)}{8 c_w s_w}\\
\widetilde{f}_2^{AAZ}&=& \frac{ \left(\text{q}_1^2+\text{q}_2^2\right)
   v^2 \left(s_w{}^2 \frac{C_{WW}}{\Lambda ^4}-2 c_w s_w \frac{C_{BW}}{\Lambda ^4}+4 c_w{}^2 \frac{C_{BB}}{\Lambda
   ^4}\right)}{16 c_w s_w}\\
   \widetilde{f}_3^{AAZ}&=& \frac{ \left(\text{q}_1^2-\text{q}_22\right) v^2 \left(s_w{}^2 \frac{C_{WW}}{\Lambda ^4}-2 c_w s_w
   \frac{C_{BW}}{\Lambda ^4}+4 c_w{}^2 \frac{C_{BB}}{\Lambda ^4}\right)}{16 c_w s_w}\\
    \widetilde{f}_4^{AAZ}&=&0,
\end{eqnarray}

\begin{eqnarray}
\widetilde{f}_1^{ZZA}&=& 0\\
\widetilde{f}_2^{ZZA}&=& \frac{ \text{q}_3^2 v^2 \left(-c_w s_w
  \frac{C_{WW}}{\Lambda ^4}+\frac{C_{BW}}{\Lambda ^4} \left(c_w{}^2-s_w{}^2\right)+4 c_w s_w \frac{C_{BB}}{\Lambda
   ^4}\right)}{8 c_w s_w}\\
   \widetilde{f}_3^{ZZA}&=& \frac{ \left(\text{q}_1^2-\text{q}_2^2\right) v^2 \left(-c_w s_w \frac{C_{WW}}{\Lambda
   ^4}+\frac{C_{BW}}{\Lambda ^4} \left(c_w{}^2-s_w{}^2\right)+4 c_w s_w \frac{C_{BB}}{\Lambda ^4}\right)}{8 c_w
   s_w}\\
    \widetilde{f}_4^{ZZA}&=&0
\end{eqnarray}
and
\begin{eqnarray}
\widetilde{f}_1^{ZZZ}&=& -\frac{
   \left(\text{q}_1^2-\text{q}_2^2\right) v^2 \left(c_w{}^2 \frac{C_{WW}}{\Lambda ^4}+2 c_w s_w \frac{C_{BW}}{\Lambda
   ^4}+4 s_w{}^2 \frac{C_{BB}}{\Lambda ^4}\right)}{8 c_w s_w}\\
\widetilde{f}_2^{ZZZ}&=& -\frac{ \left(\text{q}_1^2+2
   \text{q}_1.\text{q}_2\right) v^2 \left(c_w{}^2 \frac{C_{WW}}{\Lambda ^4}+2 c_w s_w \frac{C_{BW}}{\Lambda ^4}+4 s_w{}^2
   \frac{C_{BB}}{\Lambda ^4}\right)}{8 c_w s_w}\\
   \widetilde{f}_3^{ZZZ}&=& \frac{ \left(2 \text{q}_1.\text{q}_2+\text{q}_2^2\right) v^2 \left(c_w{}^2 \frac{C_{WW}}{\Lambda ^4}+2 c_w s_w
   \frac{C_{BW}}{\Lambda ^4}+4 s_w{}^2 \frac{C_{BB}}{\Lambda ^4}\right)}{8 c_w s_w}
\end{eqnarray}
for the CP-violating couplings. We have dropped the star in the exponent of the coupling and the dependence on the momenta to shorten the notation. The $\text{q}_i$ are the momenta of the bosons as defined in \cite{Gounaris:2000dn}. We do not give the $g$ couplings since we assume that the bosons are on-shell or connected to a massless fermion line. 
Those couplings have been translated to the parametrization of Eqs.~\eqref{eq:anoZZ} and \eqref{eq:anoZA} using the relation from ref.~\cite{Gounaris:2000dn}.

\bibliography{biblio}
\bibliographystyle{JHEP}

\end{document}